\documentclass[aps,prl,reprint,groupedaddress]{revtex4-1}
\usepackage{bm}
\usepackage{graphicx}
\usepackage[usenames]{color}
\usepackage{amsmath,amsfonts,amssymb}
\usepackage{hyperref}
\usepackage{ulem}

\newcommand{\ket}[1]{|{#1}\rangle}

\begin{document}
\title{Higher-order Topological Mott Insulators}

\author{Koji Kudo$^1$}
\author{Tsuneya Yoshida$^{1,2}$}
\author{Yasuhiro Hatsugai$^{1,2}$}
\affiliation{
$^1$Graduate School of Pure and Applied Sciences, University of Tsukuba, 
Tsukuba, Ibaraki 305-8571, Japan\\
$^2$Department of Physics,
University of Tsukuba, Tsukuba, Ibaraki 305-8571, Japan
}

\date{\today}

\begin{abstract}
 We propose a new correlated topological state which we call a higher-order topological Mott insulator (HOTMI). This state exhibits a striking bulk-boundary correspondence due to electron correlations.
 Namely, the topological properties in the bulk, characterized by the $\mathbb{Z}_3$ spin-Berry phase, result in gapless corner modes emerging only in
spin excitations (i.e., the single-particle excitations remain gapped around the corner). 
We demonstrate the emergence of the HOTMI in a Hubbard model on the kagome lattice, and elucidate how strong correlations change gapless corner modes at the noninteracting case.

\end{abstract}
\maketitle

\paragraph{Introduction.--}
The discovery of topological insulators/superconductors has opened a new field of study in condensed matter physics~\cite{TKNN,Kane_PRL05,Kane_Review,Qi_Review,Konig_HgTe_Science07}. 
A remarkable phenomenon of these states is the bulk-edge correspondence~\cite{Hatsugai_BEC_PRL93,Hatsugai_BEC_PRB93}; because of the topological properties in the $d$-dimensional bulk, gapless edge states emerge around $d-1$-dimensional boundaries of the system which result in the quantized electron-magnetic responses~\cite{TKNN,Qi_RPB08} and the emergence of Majorana fermions~\cite{Kitaev_Majorana_IOP01,Alicia_Majorana_IOP12,Sato_Majorana_JPSJ16,Mourik_Majorana_Science12,Rokhinson_Majorana_NatPhys12,Das_Majorana_NatPhys12}.
So far impacts of correlation effects on topological states 
have been addressed as
one of the significant issues of this field~\cite{Hohenadler_TI+U_PRL11,Yu_TI+U_PRL11,Yoshida_TIvsMI_PRB12,Tada_TI+U_PRB12,Yoshida_TIvsAFMIPRB13,Hohenadler_TI+U_review,Rachel_TI+U_review}. 
As the results of extensive studies, a variety of new phenomena have been reported which are induced by correlation effects on the gapless boundary states. For example, electron correlations may change the topological classification~\cite{Fidkowski_ZtoZ8_RPB10_1,Fidkowski_ZtoZ8_RPB11,Turner_ZtoZ8_RPB11,Lu_ZtoZn_CS_PRB12,Levin_ZtoZn_CS_PRB12,Yao_ZtoZn_PRB13,Ryu_ZtoZn_PRB13,You_ZtoZn_PRB14,Isobe_ZtoZ4_PRB14,Yoshida_ZtoZ4_PRB15,Morimoto_ZtoZn_PRB15,Yoshida_ZtoZ4_PRB17,Yoshida_ZtoZ8_2D_PRL17,Yoshida_ZtoZ4_1D_PRL18} which plays an important role for material searching. 
Furthermore, correlation effects induce the so-called topological Mott insulating state where systems show Mott physics only around the edges with maintaining nontrivial properties in the bulk~\cite{Pesin_TMI_NatPhys10,PhysRevB.82.075106,Yoshida_TMI_1D_PRL14,Yoshida_TMI_2D_PRB16,Kargarian_TCMI_PRL13,foot_note_TMI}.

Along with the above significant progress of correlated topological phases, new classes of topological insulators/superconductors have been proposed which are referred to as higher-order topological insulators/superconductors~\cite{Hatsugai_HOTI_EPL11,Hashimoto_HOTI_PRB17,Benalcazar_HOTI_Science17,Schindler_HOTI_SciAdv,Benalcazar_HOTI_PRB17,Hayashi_HOTI_MathPhys18,HOTI_Thomale18,Araki_HOTI_PRB18,HOTI_Ghorashi19}. 
These phases exhibit the characteristic bulk-boundary correspondence; topological properties in the $d$-dimensional bulk predict gapless charge excitations around $d-2$- or $d-3$-dimensional boundaries, while gapless charge excitations are absent around $d-1$-dimensional boundaries. 
Recent theoretical and experimental studies elucidate that the above higher-order bulk-edge correspondence is ubiquitous phenomenon; it can be observed for a variety of materials~\cite{Schindler_hingeBi_NatPhys18,Garcia_photonic_hinge_numerical_Nat18,Yan_Majorana_corner_PRL18,CYue_NatPhys19}.
For instance, hinge states are observed for a hexagonal pit on a bismuth (111) surface~\cite{Schindler_hingeBi_NatPhys18}.
In addition, Majorana corner states are theoretically proposed for an ordinary two-dimensional topological insulator proximity to a cuprate superconductor~\cite{Yan_Majorana_corner_PRL18}.

The above recent progresses lead us the following crucial question: \textit{what is the impact of electron correlations to higher-order topological phases?} Correlation effects of boundary modes have recently addressed based on a field theory~\cite{You_INTHOTI_PRB18}. 
We would like to stress however that systematic analysis both for the bulk and boundaries may elucidate new topological states which is significant but still missing.

In this paper, we explore correlated systems from the above higher-order perspective.
Our systematic analysis both for the bulk and boundaries discovers a novel topological state, \textit{a higher-order topological Mott insulator} (HOTMI), which exhibits a novel bulk-boundary correspondence due to electron correlations. Namely, in contrast to the above mentioned higher-order topological insulators (HOTIs) for free-fermions, the topological properties in the $d$-dimensional bulk do not induce the gapless charge excitations. Instead, the bulk nontrivial properties result in gapless spin excitations around the $d-2$- or $d-3$-dimensional boundaries.
Our analysis based on the exact diagonalization~\cite{nizigen_sa-ti}
verifies the emergence of HOTMIs in a Hubbard model on the kagome lattice. Specifically, our systematic numerical simulation elucidates that the system possesses topologically nontrivial properties characterized with the $\mathbb{Z}_3$
spin-Berry phase. Correspondingly, the system hosts gapless spin excitations around the corner while correlations effects destroy the gapless charge excitations. The above behaviors are also confirmed by analyzing an effective Hamiltonian for the strongly correlated limit.

\begin{figure}[t]
 \begin{center}
 \includegraphics[width=\columnwidth]{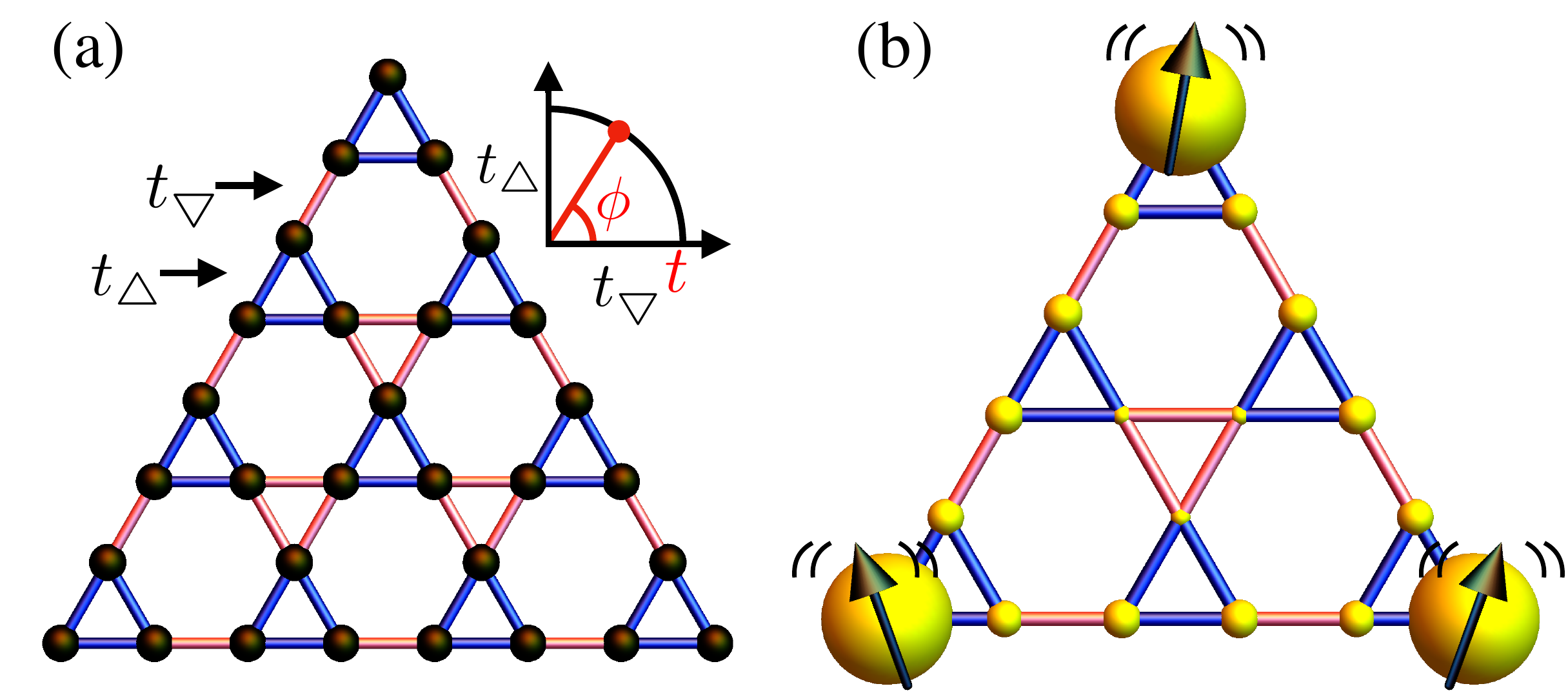}
 \end{center}
 \caption{
 (a) Sketch of the kagome lattice with $N_\text{UC}=10$ and
 the boundary condition. Blue and red bonds represent nearest-neighbor hoppings
 corresponding to the Hamiltonian $H_\text{kin}^\bigtriangleup$ and 
 $H_\text{kin}^\bigtriangledown$, respectively.
 (b) $\langle\bm{S}_i^2\rangle-\langle\bm{S}_i^2\rangle_0$ computed by the Lanczos method for $N_\text{UC}=6$, $t_\bigtriangleup/t_\bigtriangledown=0.4$.
 In panel (b), the calculated value is represented by the radius of the spheres which is obtained under the OBC.
 The spin operator can be rewritten as $\bm{S}^2_i=3(n_{i\uparrow}+n_{i\downarrow}-2n_{i\uparrow}n_{i\downarrow})/4$.
 }
 \label{fig:S_SQUARE}
\end{figure}

\paragraph{Model and topological invariant.--} 
Let us begin by describing our model and introducing
the $\mathbb{Z}_3$ spin-Berry phase.
We consider the system of spinful interacting electrons on the kagome
lattice~\cite{Imai_KGM_PRB03,Ohashi_KGM_PRL06,Udagawa_KGM_PRL10,Furukawa_KGM_PRB10,Yamada_KGM_PRB11,Guertler_KGM_PRB14,Kim_KGM_PRB15,Lee_KGM_PRB_18,Nakamura_KGM_EPJB18}.
The Hamiltonian is given by 
$H= H^\bigtriangleup_\text{kin}+H^\bigtriangledown_\text{kin}+H_\text{int},$
\begin{subequations}
\label{Eq:Hami_ele}
with
\begin{eqnarray}
 \label{Eq:ham1}
 H_\text{kin}^\gamma 
 &=&t_\gamma \sum_{i,j\in\gamma}
 \sum_{\alpha,\beta=\uparrow,\downarrow}
 c^\dagger_{i\alpha}\sigma^z_{\alpha\beta}c_{j\beta}
 +\text{h.c.},\\
 \label{Eq:ham2}
 H_\text{int}
 &=&
 U\sum_{i}\left(n_{i\uparrow}-\frac{1}{2}\right)\left(n_{i\downarrow}-\frac{1}{2}\right).
\end{eqnarray}
\end{subequations}
Here, $\gamma=\bigtriangleup$ or $\bigtriangledown$,
$n_{i\alpha}=c_{i\alpha}^\dagger c_{i\alpha}$, $c_{i\alpha}^\dagger$
($c_{i\alpha}$) is the creation (annihilation) operator on site $i$ with
spin $\alpha=\uparrow$ or $\downarrow$, and
$i,j\in\bigtriangleup(\bigtriangledown)$ indicates the summation over the
nearest-neighbor pairs $i<j$ belonging to upward (downward) triangles,
see Fig.~\ref{fig:S_SQUARE}(a).
We parametrize the hopping parameters as $t_\bigtriangleup=t\sin{\phi}$ and 
$t_\bigtriangledown=t\cos{\phi}$ with $0\leq\phi\leq\pi/2$, and suppose the 
relations $0\leq U $ and $0<t$.
Since the phase of the hopping is spin-dependent due to the Pauli matrix 
$\sigma^z$ in Eq.~\eqref{Eq:ham1}, time-reversal symmetry is 
broken while the system preserves U(1) spin-rotational symmetry. In the 
following, we focus on the system at 
half-filling~\cite{PHS}.
We note that the spin-dependent hoppings 
play an essential role
in realizing the gapped phase for the half-filled noninteracting system.
In Fig.~\ref{fig:S_SQUARE}(a), we show a sketch of the system with $N_\text{UC}=10$ under the open boundary condition (OBC), where $N_\text{UC}$ is the number of the unit cells.

To characterize the bulk topological properties, we introduce the spin-Berry phase that is the spin counterpart of the $\mathbb{Z}_3$ topological invariant~\cite{Hatsugai_HOTI_EPL11,Kawarabayashi_KGM_JPSJ19}.
Firstly, we define the spin-Berry connection whose integral corresponds to $\mathbb{Z}_3$ spin-Berry phase.
The spin-Berry connection is defined as $A_-(\vec{\theta})=\langle G_-(\vec{\theta})|dG_-(\vec{\theta})\rangle$. 
Here, $|G_-(\vec{\theta})\rangle$ denotes the ground state of the modified Hamiltonian by a so-called local gauge twist which is defined as follows. 
We pick up a specific downward triangle, which includes the sites $1$, $2$, $3$, and rewrite the Hamiltonian
$H_\text{kin}^{\bigtriangledown}$ 
as
$U_-(\vec{\theta})H_\text{kin}^{\bigtriangledown}U_-^\dagger(\vec{\theta})$ with  $\vec{\theta}:=(\theta_1,\theta_2)$, $U_-(\vec{\theta})=e^{in_1^-\theta_1}e^{-in_2^-\theta_2}$, and
$n_j^-=n_{j\uparrow}- n_{j\downarrow}$. 
With the spin-Berry connection $A_-$, the $\mathbb{Z}_3$ spin-Berry phase is defined as
$\gamma_-=\frac{1}{i}\int_CA_-.$
Here, $C$ represents the following path of the integral: $\vec{\theta}=\vec{f}(t)$ ($0\leq t\leq1$) with $\vec{f}(t)=2\pi(t,t)$ for $0\leq t<1/3$ and $\vec{f}(t)=2\pi(t,1/2-t/2)$ for $1/3\leq t\leq1$.
To numerically compute the integral, we apply the method proposed in Ref.~\onlinecite{Fukui_FHW_JPSJ06}.
This topological invariant is well-defined
even if the system is
interacting. The $C_3$ symmetry in the system brings its quantization as
$\gamma_-=2\pi n/3$ with $n=0,1,2$.

Let us briefly review the behaviors of noninteracting case~\cite{Hatsugai_HOTI_EPL11,Ezawa_HOTI_KGM_PRL18}.
For $U/t=0$, our model provides three phases: (i) a HOTI phase with $\gamma_-=2\pi/3$ for $0\leq t_\bigtriangleup/t_\bigtriangledown<1/2$,  (ii) a metallic phase for $1/2\leq t_\bigtriangleup/t_\bigtriangledown\leq2$, and (iii) a trivial phase with $\gamma_-=0$ for $2<t_\bigtriangleup/t_\bigtriangledown<\infty$. 
The topological properties in the bulk for phase (i) result in gapless excitations for the single-particle spectrum around the corners.
The average magnetization per unit cell is one-half in the gapped phases, (i) and (iii),
which arises from the absence of time-reversal symmetry (for more details, see Sec.~\ref{Sec:NONINT_HOTI} of  Supplemental Material~\cite{sup}). 

\paragraph{Overview of numerical results.--}
Before entering upon detailed discussions, let us overview our numerical results.

The interplay between correlation effects and the topological properties induces the HOTMI exhibiting a novel bulk-boundary correspondence.
In contrast to the noninteracting case, the topological properties are not reflected in the single-particle excitations; introducing the Hubbard interaction destroys gapless excitations for the single-particle spectrum. Instead, the bulk-nontrivial properties result in the gapless spin excitations around the corners.
Figure~\ref{fig:S_SQUARE}(b) supports the above behaviors. This figure shows the spin expectation value $\langle\bm{S}_i^2\rangle-\langle\bm{S}_i^2\rangle_0$~\cite{SHIBA_S2_PRB72} at each site under the OBC for $t_\bigtriangleup/t_\bigtriangledown=0.4$ and $N_\text{UC}=6$.
Here, $\langle \cdot \rangle$ and $\langle \cdot \rangle_0$ refer to the expectation values with respect to the ground state multiplet for $U/t=1$ and $U/t=0$, respectively. 
In Fig.~\ref{fig:S_SQUARE}(b), we can see that localized free-spins emerge only around the corners. This figure indicates the presence of gapless corner modes in the spin excitation spectrum due to the Mott behaviors occurring only around the corners.

In the following, we systematically analyze the system for the bulk and boundaries in order to verify the HOTMIs for the kagome Hubbard model.

\paragraph{Strong correlation limit.--}
As a first step, we analyze the kagome-Hubbard model in a strongly correlated limit $t/U\ll 1$.
In this limit, electrons are completely localized in the entire system. However, we can still discuss the presence/absence of gapless spin excitations around the corners which are induced by the bulk-topological properties.

Based on the standard perturbation analysis for degenerated systems, we obtain the following effective spin model (see Sec.~\ref{Sec:THIRD_ORDER} of Supplemental Material~\cite{sup}),
$H_\text{eff}= H^{(2)}_\bigtriangleup+H^{(2)}_\bigtriangledown+H^{(3)}_\bigtriangleup+H^{(3)}_\bigtriangledown$,
\begin{subequations}
\label{Eq: Hspin}
with 
\begin{eqnarray}
 H^{(2)}_\gamma
 &=&J^{(2)}_\gamma\sum_{i,j,k\in\gamma}
 \left(-\frac{1}{2}\bm{S}_{ijk}^2+S_{ijk}^{z\ 2}+\frac{3}{8}\right),
 \label{Eq:second}\\
 H^{(3)}_\gamma
 &=&J^{(3)}_\gamma\sum_{i,j,k\in\gamma}
 S_{ijk}^z\left(-3\bm{S}_{ijk}^2+4S_{ijk}^{z\ 2}+\frac{9}{4}\right). 
 \label{Eq:third}
\end{eqnarray}
\end{subequations}
Here, $J^{(n)}_\bigtriangleup=J^{(n)}\sin^n{\phi}$, $J^{(n)}_\bigtriangledown=J^{(n)}\cos^n{\phi}$, $J^{(n)}=4t^n/U^{n-1}$, $\bm{S}_{ijk}=\bm{S}_i+\bm{S}_j+\bm{S}_k$ and $i,j,k\in\bigtriangleup(\bigtriangledown)$ indicates the summation over three sites $i<j<k$ composing upward (downward) triangles. 
$H^{(2)}_\gamma$ and $H^{(3)}_\gamma$ correspond to the second- and third-order-term, respectively.
In the absence of the third-order terms, the system is reduced to a spin system of the XXZ-interaction. The XXZ model
preserves
the time-reversal symmetry while the original Hamiltonian $H$ does not. In order to remove the artificially produced symmetry, the third-order perturbation is necessary.

\begin{figure}[t]
 \begin{center}
  \includegraphics[width=\columnwidth]{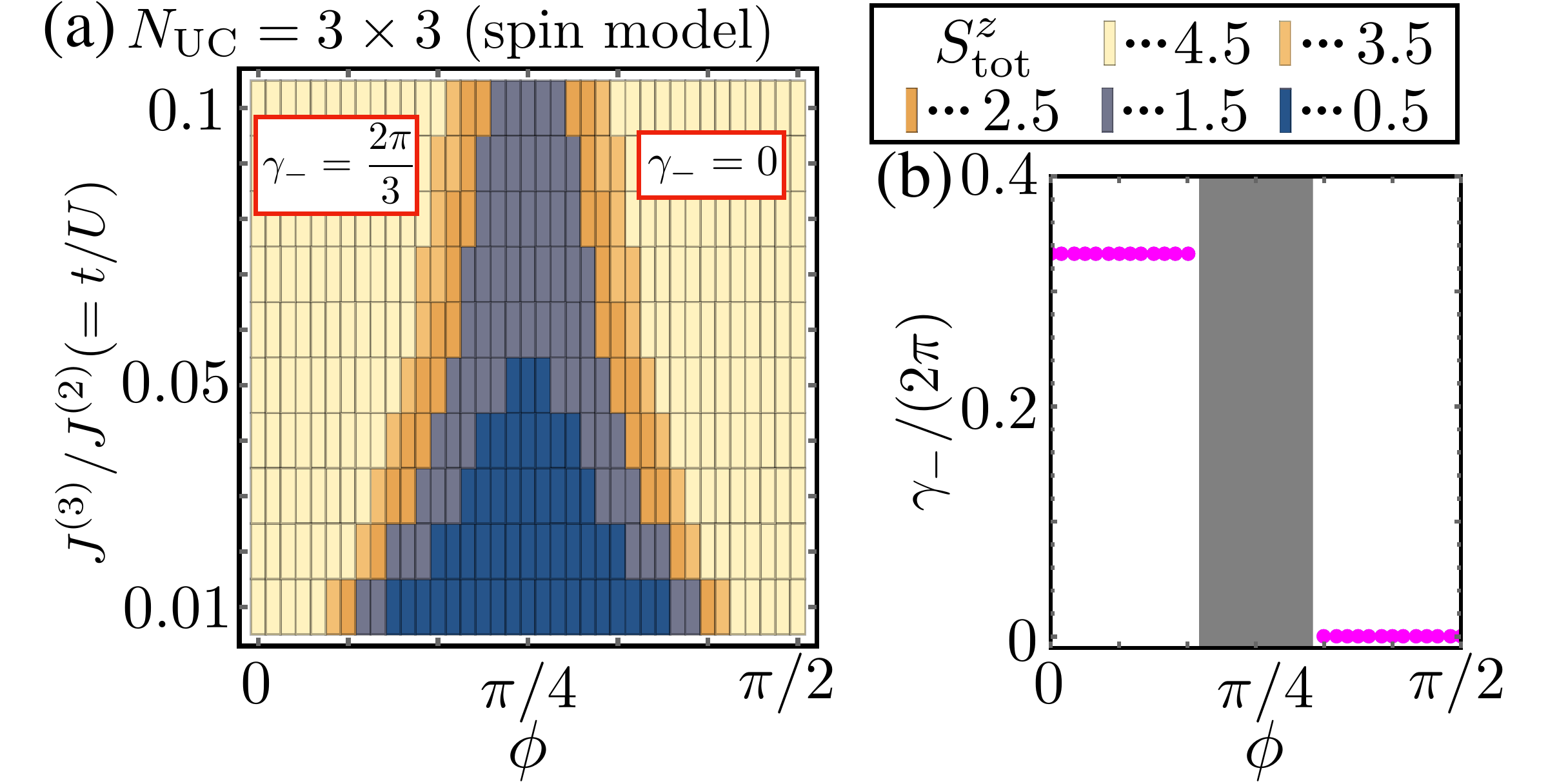}
 \end{center}
 \caption{
 Numerical results for the spin model~\eqref{Eq: Hspin}.
 (a) $S^z_\text{tot}$ of the ground state as a function of $\phi$ and
 $J^{(3)}/J^{(2)}$.
 (b) Spin-Berry phase $\gamma_-$ of the ground state for $J^{(3)}/J^{(2)}=0.1$ 
 as a  function of $\phi$. The gray region in panel (b)
 indicates that
 $S^z_\text{tot}$ of the ground state is not $N_\text{UC}/2=4.5$. The system is
 periodic and its size is $N_\text{UC}=3\times3$ for panels (a) and (b).}
 \label{fig:SPIN_PERIO}
\end{figure}
The presence of $H^{(3)}_\gamma$ is also essential for the bulk gap. To demonstrate it, let us analyze the cases for $\phi=0$ or $\pi/2$ where the system is reduced to three-site problems
under the periodic boundary condition (PBC).
When the effective model includes only $H^{(2)}_\gamma$, the Kramers pair $|S_{123}=3/2,S_{123}^z=\pm1/2\rangle$ is realized as the ground state on three sites, which results in the macroscopic degeneracy in the bulk system.
Introducing $H^{(3)}_\gamma$ breaks the time-reversal symmetry and lifts this Kramers degeneracy, which result in the unique ground state $|S_{123}=3/2,S_{123}^z=1/2\rangle$.
This concise analysis suggests the emergence of the bulk gap even if the system is not completely decoupled ($0<\phi<\pi/2$) unless the energy gap associated with the above three-site problem is vanishing.
Figure~\ref{fig:SPIN_PERIO}(a) supports the presence of the bulk gap. 
In this figure, the ground state magnetization 
$S^z_\text{tot}$ 
is plotted against $\phi$ and $J^{(3)}/J^{(2)}=t/U$. We remind that U(1) spin-rotational symmetry is preserved allowing us to label the energy eigenstates with $S^z_\text{tot}$.
Figure~\ref{fig:SPIN_PERIO}(a) shows that the ground state with $S^z_\text{tot}/ N_\text{UC}=1 /2$ extends to a finite range of $\phi$, indicating that the energy gap observed for three-site problem remains finite.
The energy eigenvalues as a function of $\phi$ is plotted in Sec.~\ref{Sec:PERIO_SPIN} of Supplemental Material~\cite{sup} which directly supports the presence of the bulk gap.

Based on the result that the gap is finite in the region of $S^z_\text{tot}=4.5$ [see Fig.~\ref{fig:SPIN_PERIO}(a)], let us discuss the topological properties of the ground state with $S^z_\text{tot}=N_\text{UC}/2$.
Figure~\ref{fig:SPIN_PERIO}(b) plots its $\gamma_-$ for $J^{(3)}/J^{(2)}=0.1$ as a function of $\phi$.
The result indicates that the gapped phase for $0\leq\phi\lesssim\pi/4$ is characterized with $\gamma_-=2\pi/3$ while the one for 
$\pi/4\lesssim\phi\leq\pi/2$ 
is characterized with $\gamma_-=0$. Because the gap remains finite in the yellow-colored regions in Fig.~\ref{fig:SPIN_PERIO}(a), we can characterize the topology of these gapped phases as summarized in this figure.
We note here that the local gauge twist $U_-(\vec{\theta})H_\text{kin}^{\bigtriangledown} U_-^\dagger(\vec{\theta})$ defined in the Hubbard model is modified as
$U_\text{eff}(\vec{\theta})
(H^{(2)}_{\bigtriangledown}+H^{(3)}_{\bigtriangledown})
U_\text{eff}^\dagger(\vec{\theta})$ for the effective model, where
$U_\text{eff}=e^{2iS^z_1\theta_1}e^{-2iS^z_2\theta_2}$.
The $\mathbb{Z}_3$ spin-Berry phase predicts to which decoupled limit
($\phi=0,\pi/2$) a gapped phase is adiabatically connected.

So far, we have elucidated the topological properties in the bulk. In order to verify a novel bulk-boundary correspondence, we here analyze the systems under the OBC [Fig.~\ref{fig:S_SQUARE}(a)].
In a similar way as the previous case, we start with the cases of $\phi=0$ and $\pi/2$ where the system is reduced to three-site problems.
At $\phi=0$, the coupling for upward triangle is zero ($H^{(2)}_\bigtriangleup=H^{(3)}_\bigtriangleup=0$). 
Besides the trimer in the bulk, we can see the dimers for the one-dimensional edges due to the XXZ-spin interaction ($H^{(2)}_\bigtriangledown$).
Furthermore, we can observe three free-spins on corners which are clearly gapless. 
These free-spins result in the eight-fold degeneracy of the ground state for the OBC. 
At $\phi=\pi/2$, the coupling for downward triangle is zero
($H^{(2)}_\bigtriangledown=H^{(3)}_\bigtriangledown=0$). In this case, the
system is composed only of trimers so that the ground state is unique
and its $S^z_\text{tot}$ is given as $S_\text{tot}^z=N_\text{UC}/2$.

\begin{figure}[t]
 \begin{center}
  \includegraphics[width=\columnwidth]{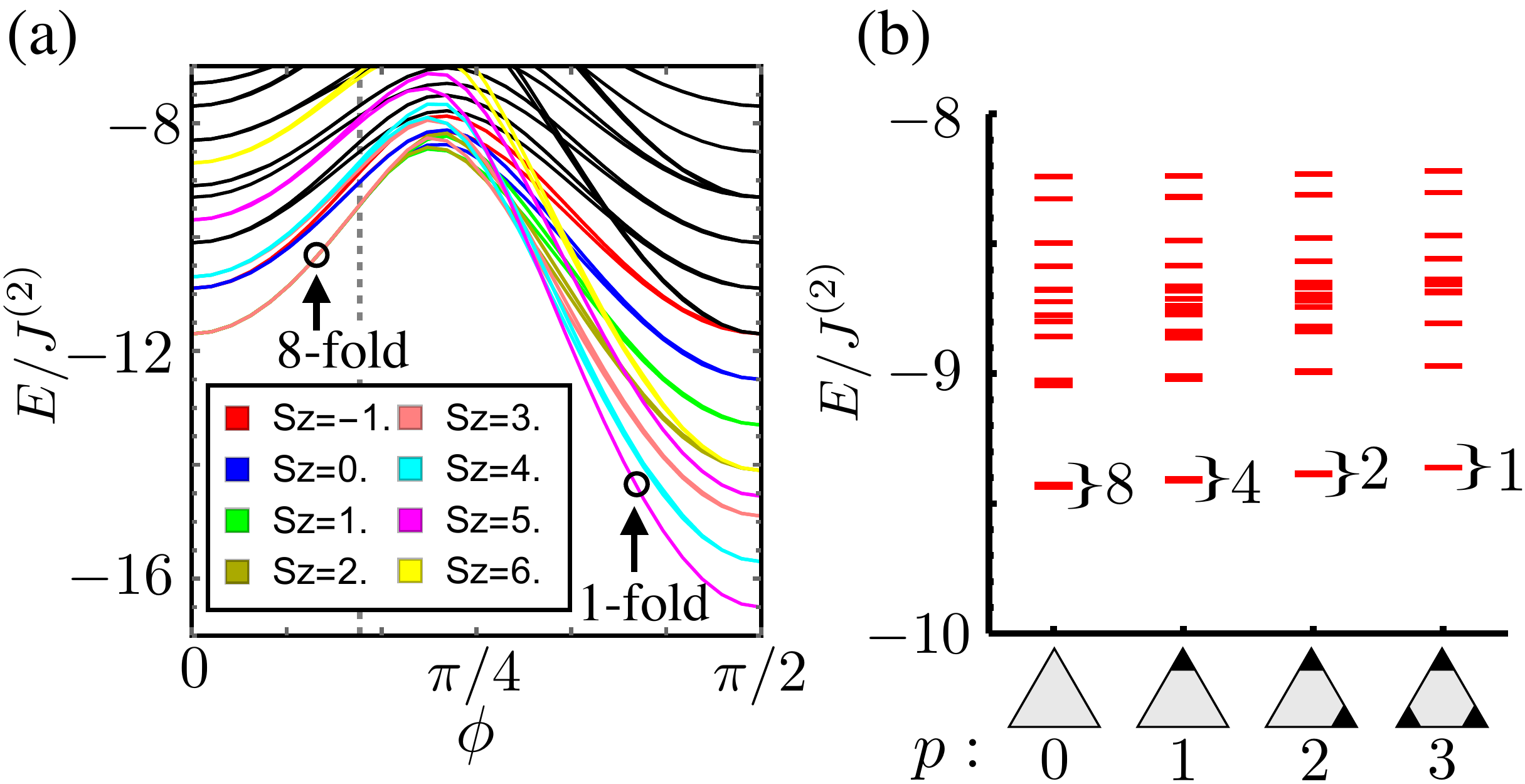}
 \end{center}
 \caption{
 Numerical results for the spin model~\eqref{Eq: Hspin}.  
 (a) Energy spectra $E/J^{(2)}$ as functions of $\phi$.
 (b) Energy spectra $E/J^{(2)}$ at $\tan{\phi}=0.5$ for each value of $p$,
 where $p$ is the number of removed sites.
 Panels (a) and (b) are obtained for $N_\text{UC}=10$ and $J^{(3)}/J^{(2)}=0.1$
 under the OBC. Only four lowest energies in the subspace specified by
 $S^z_\text{tot}$ are shown here. The black lines in panel (a)
 express the energy for $S^z_\text{tot}<-1$ or $6<S^z_\text{tot}$. 
 The dashed line in panel (a) represents $\tan{\phi}=0.5$.
 }
 \label{fig:SPIN_OPEN}
\end{figure}
To demonstrate the existence of this bulk-boundary correspondence at the
other $\phi$, we show in Fig.~\ref{fig:SPIN_OPEN}(a) the $\phi$ dependence of 
the energy $E/J^{(2)}$ for $J^{(3)}/J^{(2)}=0.1$
under the OBC with $N_\text{UC}=10$ [see, 
Fig.~\ref{fig:S_SQUARE}(a)]. 
For $0\leq\phi\lesssim\pi/4$ ($\pi/4\lesssim\phi\leq\pi/2$), the eight-fold
degenerate (unique) ground state is observed.
Their $S^z_\text{tot}$ ranges between 0 and 3
(see Sec.~\ref{Sec:spin_Sz} of Supplemental Material~\cite{sup}).
To demonstrate that the 8-fold degeneracy is attributed to the gapless corner 
states, we remove a site located at each of ends
(corners) and examine its effects on the degeneracy.
Figure~\ref{fig:SPIN_OPEN}(b) plots the energy spectra for the four types of the
geometry. Clearly, the ground state has $2^{3-p}$-fold degeneracy corresponding
to the defects of the $p$ ends (corners), 
which suggests the emergence of the gapless modes at the $3-p$ corners
for $0\leq\phi\lesssim\pi/4$. 
Consequently, these results indicate that 
the HOTMI state emerges in the region with $\gamma_-=2\pi/3$ shown in Fig.~\ref{fig:SPIN_PERIO}(a) which exhibits gapless corner states appearing only in spin excitations.

In the above, by analyzing the effective spin model, we have elucidated that the system exhibits the bulk-boundary correspondence of the HOTMI. 
Let us next analyze the Hubbard model beyond the framework of this effective theory.

\paragraph{From weakly to strongly correlated regions.--}
In response to the results of the effective spin model, two questions arise. 
``Does the HOTMI emerge in the Hubbard model even for the finite $U/t$?''
``If so, how the noninteracting HOTI state changes into the HOTMI state by 
electron correlations?''
In the rest of this work, we address these issues by diagonalizing the Hamiltonian~(\ref{Eq:Hami_ele}) and provide a numerical evidence that the HOTMI is exhibited in the system as long as $U/t\neq0$.

\begin{figure}[t]
 \begin{center}
  \includegraphics[width=\columnwidth]{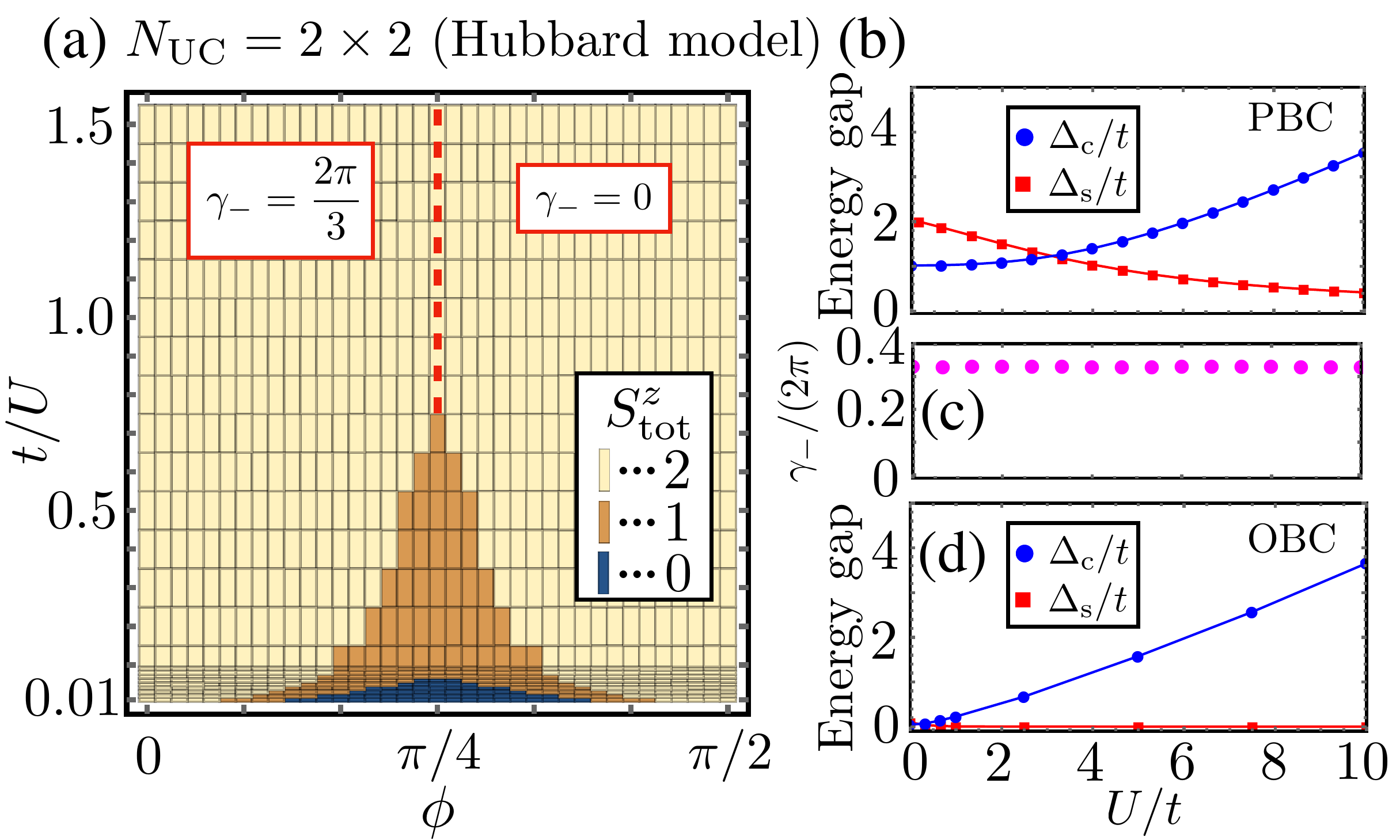}
 \end{center}
 \caption{
 Numerical results for the Hubbard model~\eqref{Eq:Hami_ele}.
 (a) $S^z_\text{tot}$ of the ground state as a function of $\phi$ and $t/U$.
 (b) Charge gap $\Delta_\text{c}$ and spin gap $\Delta_\text{s}$ as functions 
 of $U/t$. (c) Spin-Berry phase $\gamma_-$ as functions of $U/t$.
 (d) Charge gap $\Delta_\text{c}$ and spin gap $\Delta_\text{s}$ as functions
 of $U/t$.
 Panels (a), (b), and (c) are obtained for $N_\text{UC}=2\times2$ under
 the PBC. Panel (d) is obtained for $N_\text{UC}=6$ under the OBC.
 We set $t_\bigtriangleup/t_\bigtriangledown=0.4$ in panels (b), 
 (c), and (d).
 }
 \label{fig:HUBBARD}
\end{figure}

Let us first investigate the bulk properties.
In Fig.~\ref{fig:HUBBARD}(a), we plot ground state magnetization
$S^z_\text{tot}$
for the PBC against $\phi$ and $t/U$ in the same manner as the previous case.
As for the region with $S^z_\text{tot}/N_\text{UC}=1/2$ [i.e., the region with $S^z_\text{tot}=2$ in Fig.~\ref{fig:HUBBARD}(a)], it is suggested that the strongly and weakly correlated systems for $0\leq\phi\lesssim\pi/4$ are adiabatically  connected with each other. Indeed, the energy gaps of the HOTI survives against the electron correlations. In Fig.~\ref{fig:HUBBARD}(b), we show the $U/t$ dependence of the charge and spin gaps for $t_\bigtriangleup/t_\bigtriangledown=0.4$.
Here, they are defined as
$\Delta_c
=(E_{\mathcal{N}_\text{e}+1,\mathcal{S}_\text{tot}^z+1/2}
+E_{\mathcal{N}_\text{e}-1,\mathcal{S}_\text{tot}^z-1/2}
-2E_{\mathcal{N}_\text{e},\mathcal{S}_\text{tot}^z})/2$
and
$\Delta_s
=(E_{\mathcal{N}_\text{e},\mathcal{S}_\text{tot}^z+1}
+E_{\mathcal{N}_\text{e},\mathcal{S}_\text{tot}^z-1}
-2E_{\mathcal{N}_\text{e},\mathcal{S}_\text{tot}^z})/2$,
where
$\mathcal{N}_\text{e}$ and $\mathcal{S}_\text{tot}^z$ is the number of 
electrons and $S^z_\text{tot}$ of the ground state, respectively 
[$\mathcal{N}_\text{e}=3\times N_\text{UC}=12$ and 
$\mathcal{S}_\text{tot}^z=2$ in Fig.~\ref{fig:HUBBARD}(b)].
Figure~\ref{fig:HUBBARD}(b) indicates that $\Delta_c$ becomes of the 
order of $U$ for large interactions, which clearly demonstrates that the HOTI 
for $U/t=0$ is adiabatically connected into the HOTMI. 
Thus, they share the same topological character, 
which is indeed confirmed in Fig.~\ref{fig:HUBBARD}(c), while they
give the different bulk-boundary correspondence.
In Fig.~\ref{fig:HUBBARD}(a), there is no sign of the metallic behavior 
observed for the noninteracting case. This is due to a finite size effect, 
see Sec.~\ref{Sec:Twist} of Supplemental Material for more details~\cite{sup}.

Let us next analyze the system under the OBC, and show that the HOTI state
expected to be transformed into the HOTMI state by the infinitesimal 
interaction
for a sufficiently large system size. Figure~\ref{fig:HUBBARD}(d) plots $\Delta_c$ and $\Delta_s$ with 
$N_\text{UC}=6$ for $t_\bigtriangleup/t_\bigtriangledown=0.4$ as functions of $U/t$. 
Here, the system gives the eight-fold degenerated ground state whose 
$S_\text{tot}^z$ ranges between -1 and 2 (see Secs.~\ref{Sec:spin_Sz} and
\ref{Sec:8fold} of 
Supplemental Material~\cite{sup}) so that 
we set $\mathcal{N}_\text{e}=18$ and $\mathcal{S}_\text{tot}^z=1$
to calculate the gaps $\Delta_c$ and $\Delta_s$. 
Figure~\ref{fig:HUBBARD}(d) implies that the gapless charge excitation 
is opened by 
the infinitesimal interaction $U$, which is consistent with the energy cost proportional to $U$ induced by
the double occupancy of a completely isolated site.
On the other hand, the spin excitations remains gapless. 
We note that the behavior of $\Delta_c$ and $\Delta_s$ is associated with a 
kind of the Mott insulators with the finite charge gap and the zero spin 
gap
exhibited at the corners. 
Thus, we call the gapless corner modes the 
corner-Mott states.
The above results elucidate the novel bulk-boundary correspondence whose microscopic origin is the emergence of the corner-Mott states observed for $0<U$. 
The bulk-boundary correspondence arising from electron correlations is precisely the defining feature of the HOTMI which is shown in Fig.~\ref{fig:S_SQUARE}(b).

Before closing, we note that the HOTMI state is consider to be ubiquitous.
For example, the extension to other two-dimensional lattice models is 
expected to be possible. 
The formulation of the $\mathbb{Z}_Q$ Berry phase for lattice models
such as kagome and pyrochlore lattices~\cite{Hatsugai_HOTI_EPL11} is recently 
generalized into any lattice model~\cite{2019arXiv190600218A};
e.g. the $\mathbb{Z}_4$ Berry phase can be defined in the square 
lattice HOTI. By combining it with our approach, one expects that the square 
HOTI state~\cite{Benalcazar_HOTI_PRB17,2019arXiv190600218A} 
would be transformed into the HOTMI state by the correlation effects.
Furthermore, the HOTMI would be extended even to three-dimensions.
This is because our approach can straightforwardly be applied to a 
system of the 
pyrochlore lattice, where its topology is characterized by $\mathbb{Z}_4$ 
spin-Berry phase~\cite{Hatsugai_HOTI_EPL11}.
Indeed, our simple numerical calculations suggest that
the correlation effect changes the pyrochlore 
HOTI state~\cite{Ezawa_HOTI_KGM_PRL18} into the HOTMI state,
which will be discussed elsewhere in details.

\paragraph{Conclusion.--}
In this paper, we discovered a novel topological state, a HOTMI which exhibits a striking bulk-boundary correspondence. 
Namely, in contrast to higher-order topological insulators for free-fermions, the topology of $d$-dimensional HOTMIs results in the gapless $d-2$- or $d-3$-dimensional boundary states 
emerging only in the spin excitation spectrum; the single-particle spectrum remains gapped even around these boundaries. 
Our exact diagonalization analysis has verified the emergence of HOTMIs in the Hubbard model on the kagome lattice. 
Specifically, we have
observed that the system possesses topological properties characterized by the spin-Berry phase with $\gamma_-=2\pi/3$. We also observed that electron correlations open the
charge gap around the corner, while the spin excitations remain gapless. These numerical results demonstrate the above novel bulk-boundary correspondence of HOTMIs.

\begin{acknowledgments}
 We thank S. Hayashi, T. Mizoguchi, and H. Araki for fruitful comments.
 K. K. thanks the Supercomputer Center, the Institute for Solid State
 Physics, the University of Tokyo for the use of the facilities.
 The work is supported by JSPS KAKENHI Grant Numbers JP16K13845 (Y.H.),
 JP17H06138, JP18H05842 (T.Y.), and JP19J12317 (K.K.).
\end{acknowledgments}

\bibliographystyle{apsrev4-1}
\bibliography{citation}

\begin{thebibliography}{73}
\bibitem [{PHS()}]{PHS}%
  \BibitemOpen
  \href@noop {} {}\bibinfo {note} {The model is also invariant under the
  transformation $c^\dagger_{i\alpha}\to\sigma^x_{\alpha\beta} c _{i\beta}$,
  which is a certain type of particle-hole symmetry.}\BibitemShut {Stop}%
\end{thebibliography}

\clearpage

\renewcommand{\thesection}{S\arabic{section}}
\renewcommand{\theequation}{S\arabic{equation}}
\renewcommand{\thefigure}{S\arabic{figure}}
\renewcommand{\thetable}{S\arabic{table}}
\setcounter{equation}{0}
\setcounter{figure}{0}
\setcounter{table}{0}
\setcounter{page}{1}
\makeatletter
\c@secnumdepth = 2
\onecolumngrid
\begin{center}
 \Large{Supplemental Material}
\end{center}
\twocolumngrid

\section{Noninteracting Bulk properties}
\label{Sec:NONINT_HOTI}
In this appendix, we discuss the bulk properties of the noninteracting HOTI in 
our models. We note that the case of spinless fermion system has been studied 
in Refs.~\onlinecite{Hatsugai_HOTI_EPL11,Ezawa_HOTI_KGM_PRL18}. 

\begin{figure}[t]
 \begin{center}
  \includegraphics[width=\columnwidth]{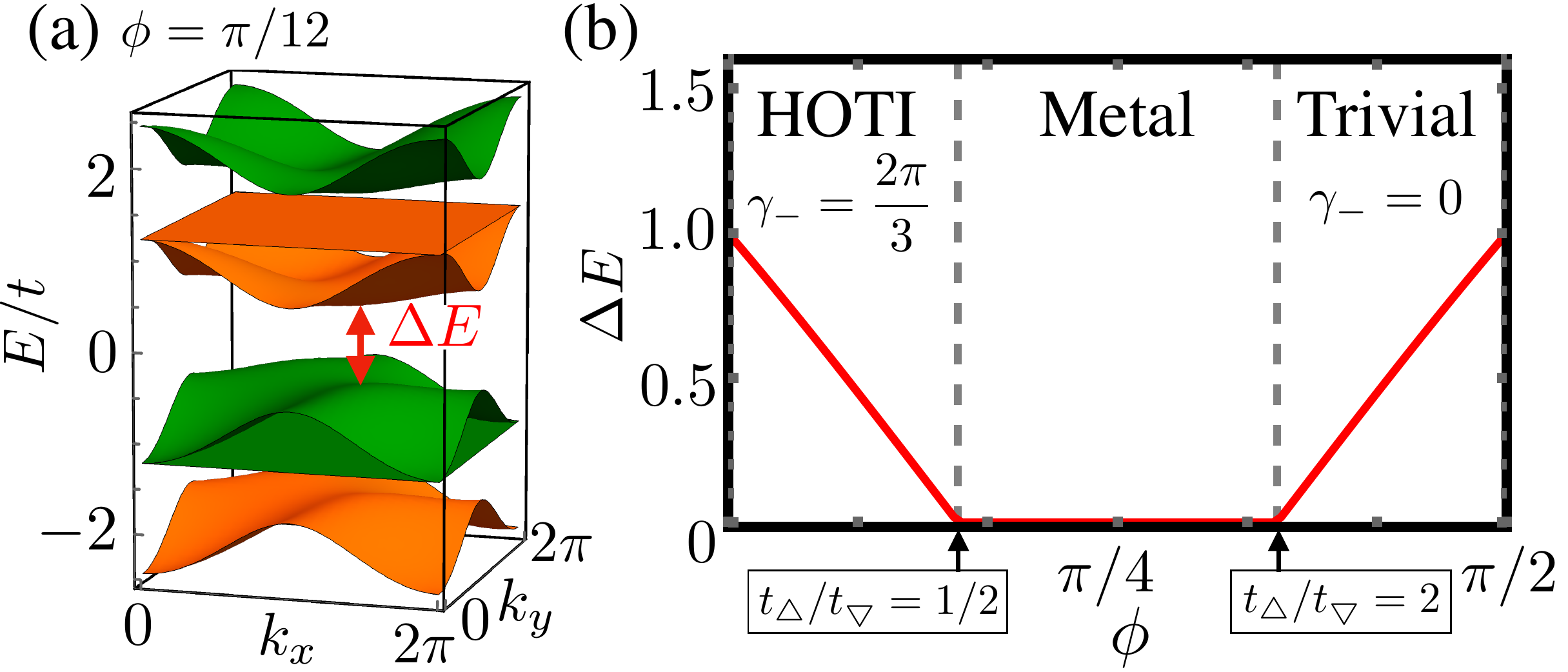}
 \end{center}
 \caption{(a) Band structure for $\phi=\pi/12$. The Green and orange bands 
 represent $S_z=1/2$ and $-1/2$, respectively. (b) Band gap $\Delta E$ between
 the third and fourth lowest bands as a function of $\phi$. Gap closing points 
 are indicated by dashed lines in panel (b).}
 \label{fig:BAND}
\end{figure}

Let us start with investigating the band structure. 
Figure~\ref{fig:BAND}(a) shows the energy bands for $\phi=\pi/12$. Due to the 
U(1) spin-rotational symmetry, we can label the bands with $S_z$. In 
Fig.~\ref{fig:BAND}(b), we plot the band gap $\Delta E$ between the third and
fourth lowest bands as a function of $\phi$.
While the states for $1/2\leq t_\bigtriangleup/t_\bigtriangledown\leq2$ are
gapless, the magnetic insulators are 
realized in both regions $t_\bigtriangleup/t_\bigtriangledown<1/2$ and 
$2<t_\bigtriangleup/t_\bigtriangledown$. The sum of $S_z$ over the
the three lowest bands is not vanishing so that the magnetization
occurs. 
As formulated in Ref.~\onlinecite{Ezawa_HOTI_KGM_PRL18}, the system for
$t_\bigtriangleup/t_\bigtriangledown<1/2$ exhibits the HOTI phase that supports
the gapless corner states under the OBC.

Next, we compute the spin-Berry phase $\gamma_-$ in the two limits $\phi=0$ and
$\pi/2$. In the system at $\phi=0$, the change of $\vec{\theta}$ is equivalent 
to the unitary transformation so that the Berry connection is given as
\begin{align}
 A_-=\langle G|U_-^\dagger dU_-|G\rangle
 =id\theta_1\langle n_1^-\rangle-id\theta_2\langle n_2^-\rangle.
\end{align}
Using $\langle n_1^-\rangle=\langle n_2^-\rangle=1/3$, we have
$\gamma_-=2\pi/3$. On the other hand, $\gamma_-$ for $\phi=\pi/2$ is vanishing 
due to the 
insensitivity of the Hamiltonian $H_\text{kin}^\bigtriangleup$ to 
$\vec{\theta}$.
Because the gap is finite for 
$t_\bigtriangleup/t_\bigtriangledown<1/2$ and
$2<t_\bigtriangleup/t_\bigtriangledown$, we can characterize the topology
of these gapped phases as summarized in Fig.~\ref{fig:BAND}(b).

\section{Third-order degenerated perturbation}
\label{Sec:THIRD_ORDER}
In this appendix, we derive the effective Hamiltonian $H^{(3)}_\gamma$ in
Eq.~\eqref{Eq:third} of the main text based on the degenerated perturbation 
theory. Let us first have a general discussion. We consider a Hamiltonian 
$H_\text{tot}=H_0+V$ with a perturbation $V$. If the ground state of the 
unperturbed problem ($H_0$ only) is degenerated and it is not lifted by the 
first-order and second-order perturbations, it is valid to diagonalize the 
effective Hamiltonian with the third-order perturbation given as
\begin{align}
 H_\text{eff}^{(3)}
 &=PV\tilde{P}\frac{1}{(E^{(0)}-H_0)^2}\tilde{P}VPVP\nonumber\\
 &+PV\tilde{P}\frac{1}{E^{(0)}-H_0}\tilde{P}V\tilde{P}\frac{1}{E^{(0)}-H_0}
 \tilde{P}VP,
\end{align}
where $P$ is the projection operator onto the degenerated ground state 
satisfying $H_0P=E^{(0)}P$, and $\tilde{P}=1-P$. 

Let us then apply the above formulation to our model. When one sets 
$H_0=H_\text{int}$ and $V=H_\text{kin}$, the projection operator $P$ is given 
by the degenerated ground state where electrons are completely localized in the
entire system. Since this
projection operator satisfies $PH_\text{int}P=0$ and $E^{(0)}=0$, we get 
\begin{align}
 H_\text{eff}^{(3)}
 =PH_\text{kin}\tilde{P}\frac{1}{H_\text{int}}\tilde{P}H_\text{kin}
 \tilde{P}\frac{1}{H_\text{int}}
 \tilde{P}H_\text{kin}P.
 \label{Eq:eff3}
\end{align}
We note that there are three $H_\text{kin}$'s in Eq.~\eqref{Eq:eff3} so that
the specific expression of 
Eq.~\eqref{Eq:eff3} is intrinsically given by solving the three-site problem.

To get the specific expression of Eq.~\eqref{Eq:eff3}, we now consider the 
three-site problem and rewrite its Hamiltonian as
\begin{subequations}
\begin{eqnarray}
 \mathcal{H}&=&\mathcal{H}_\text{kin}+\mathcal{H}_\text{int},
\end{eqnarray}
with
\begin{eqnarray}
 \mathcal{H}_\text{kin}&=&\sum_{i=1}^3\sum_{\alpha=\uparrow,\downarrow}
 t_\alpha c_{i\alpha}^\dagger c_{i+1\alpha}+\text{h.c.},\\ 
\mathcal{H}_\text{int}&=&\sum_{i=1}^3Un_{i\uparrow}n_{i\downarrow}.
\end{eqnarray}
\end{subequations}
Here, $c_{4\alpha}^\dagger=c_{1\alpha}^\dagger$. The ground state multiplet
in the unperturbed problem is given by 
\begin{align}
 \Psi
 =(
 \ket{\uparrow\uparrow\uparrow},
 \ket{\uparrow\uparrow\downarrow},
 \ket{\uparrow\downarrow\uparrow},
 \ket{\downarrow\uparrow\uparrow},
 \ket{\downarrow\downarrow\uparrow},
 \ket{\downarrow\uparrow\downarrow},
 \ket{\uparrow\downarrow\downarrow},
 \ket{\downarrow\downarrow\downarrow}
 ),
\end{align}
where $\ket{\alpha\beta\gamma}
=c_{1\alpha}^\dagger c_{2\beta}^\dagger c_{3\gamma}^\dagger\ket{0}$ and 
$\ket{0}$ is the vacuum state. The projection operators are given as 
$\mathcal{P}=\Psi\Psi^\dagger$ and $\tilde{\mathcal{P}}=1-\mathcal{P}$.
Since 
$\mathcal{H}_\text{kin}\ket{\uparrow\uparrow\uparrow}
=\mathcal{H}_\text{kin}\ket{\downarrow\downarrow\downarrow}=0$, we focus on
the sector of $S^z_{123}=\pm1/2$, where 
$\bm{S}_{123}=\bm{S}_1+\bm{S}_2+\bm{S}_3$. As an example, one can show 
\begin{align}
 \mathcal{H}_\text{eff}^{(3)}\ket{\uparrow\uparrow\downarrow}
 &=\frac{1}{U^2}\left\{
 2\left(t_\downarrow^3-t_\uparrow^3\right)
 \ket{\uparrow\uparrow\downarrow}
 +3\left(t_\uparrow^2t_\downarrow-t_\uparrow t_\downarrow^2\right)
 \ket{\uparrow\downarrow\uparrow}\right.\nonumber\\
 &\qquad\left.+3\left(t_\uparrow^2t_\downarrow-t_\uparrow t_\downarrow^2\right)
 \ket{\downarrow\uparrow\uparrow}
 \right\},
 \label{Eq:ex1}
\end{align}
where $\mathcal{H}_\text{eff}^{(3)}$ is the third-order perturbation in the 
three-site problem, see Eq.~\eqref{Eq:eff3}.
By using Eq.~\eqref{Eq:ex1}, we get
\begin{widetext}
\begin{align}
 \Psi^\dagger\mathcal{H}_\text{eff}^{(3)}\Psi
 =&\frac{1}{U^2}\left(
 \begin{array}{cccccccc}
  0 &&&&&&&\\
  &2(t_\downarrow^3-t_\uparrow^3)&3(t_\uparrow^2t_\downarrow-t_\uparrow t_\downarrow^2)&3(t_\uparrow^2t_\downarrow-t_\uparrow t_\downarrow^2)&&&&\\
  &3(t_\uparrow^2t_\downarrow-t_\uparrow t_\downarrow^2)&2(t_\downarrow^3-t_\uparrow^3)&3(t_\uparrow^2t_\downarrow-t_\uparrow t_\downarrow^2)&&&&\\
  &3(t_\uparrow^2t_\downarrow-t_\uparrow t_\downarrow^2)&3(t_\uparrow^2t_\downarrow-t_\uparrow t_\downarrow^2)&2(t_\downarrow^3-t_\uparrow^3)&&&&\\
  &&&&2(t_\uparrow^3-t_\downarrow^3)&3(t_\downarrow^2t_\uparrow-t_\downarrow t_\uparrow^2)&3(t_\downarrow^2t_\uparrow-t_\downarrow t_\uparrow^2)&\\
  &&&&3(t_\downarrow^2t_\uparrow-t_\downarrow t_\uparrow^2)&2(t_\uparrow^3-t_\downarrow^3)&3(t_\downarrow^2t_\uparrow-t_\downarrow t_\uparrow^2)&\\
  &&&&3(t_\downarrow^2t_\uparrow-t_\downarrow t_\uparrow^2)&3(t_\downarrow^2t_\uparrow-t_\downarrow t_\uparrow^2)&2(t_\uparrow^3-t_\downarrow^3)&\\
  &&&&&&&0
 \end{array}
 \right).
\end{align}
\end{widetext}
Setting the hopping parameters $t_\uparrow=-t_\downarrow=t$ in the same form
of the main text, we have
\begin{align}
 \Psi^\dagger\mathcal{H}_\text{eff}^{(3)}\Psi
 =&\frac{t^3}{U^2}\left(
 \begin{array}{cccccccc}
  0 &&&&&&&\\
  &-4&-6&-6&&&&\\
  &-6&-4&-6&&&&\\
  &-6&-6&-4&&&&\\
  &&&&4&6&6&\\
  &&&&6&4&6&\\
  &&&&6&6&4&\\
  &&&&&&&0
 \end{array}
 \right).
\end{align}
The operators related to the angular momentum coupling 
$\bm{S}_{123}$ is given as follows:
\begin{align}
 \Psi^\dagger S_{123}^z\Psi
 =\frac{1}{2}\left(
 \begin{array}{cccccccc}
  3 &&&&&&&\\
  &1&&&&&&\\
  &&1&&&&&\\
  &&&1&&&&\\
  &&&&-1&&&\\
  &&&&&-1&&\\
  &&&&&&-1&\\
  &&&&&&&-3
 \end{array}
 \right)
 \label{Eq:totSz},\\
 \Psi^\dagger\bm{S}_{123}\cdot\bm{S}_{123}\Psi
 =\frac{1}{4}\left(
  \begin{array}{cccccccc}
  15 &&&&&&&\\
  &7&4&4&&&&\\
  &4&7&4&&&&\\
  &4&4&7&&&&\\
  &&&&7&4&4&\\
  &&&&4&7&4&\\
  &&&&4&4&7&\\
  &&&&&&&15
  \end{array}
 \right).
 \label{Eq:totS2}
\end{align}
Then, using Eqs.~\eqref{Eq:totSz} and \eqref{Eq:totS2}, we get
\begin{align}
 \mathcal{H}^{(3)}_\text{eff}=J^{(3)}
 S_{123}^z
 \left(-3\bm{S}_{123}^2+4S_{123}^{z2}+\frac{9}{4}\right),
\end{align}
where $J^{(3)}=4t^3/U^2$. This formulation validates Eq.~\eqref{Eq:third} of 
the main text.

\section{Gapped topological phase in Effective spin model}
\label{Sec:PERIO_SPIN}
\begin{figure}[t]
 \begin{center}
  \includegraphics[width=\columnwidth]{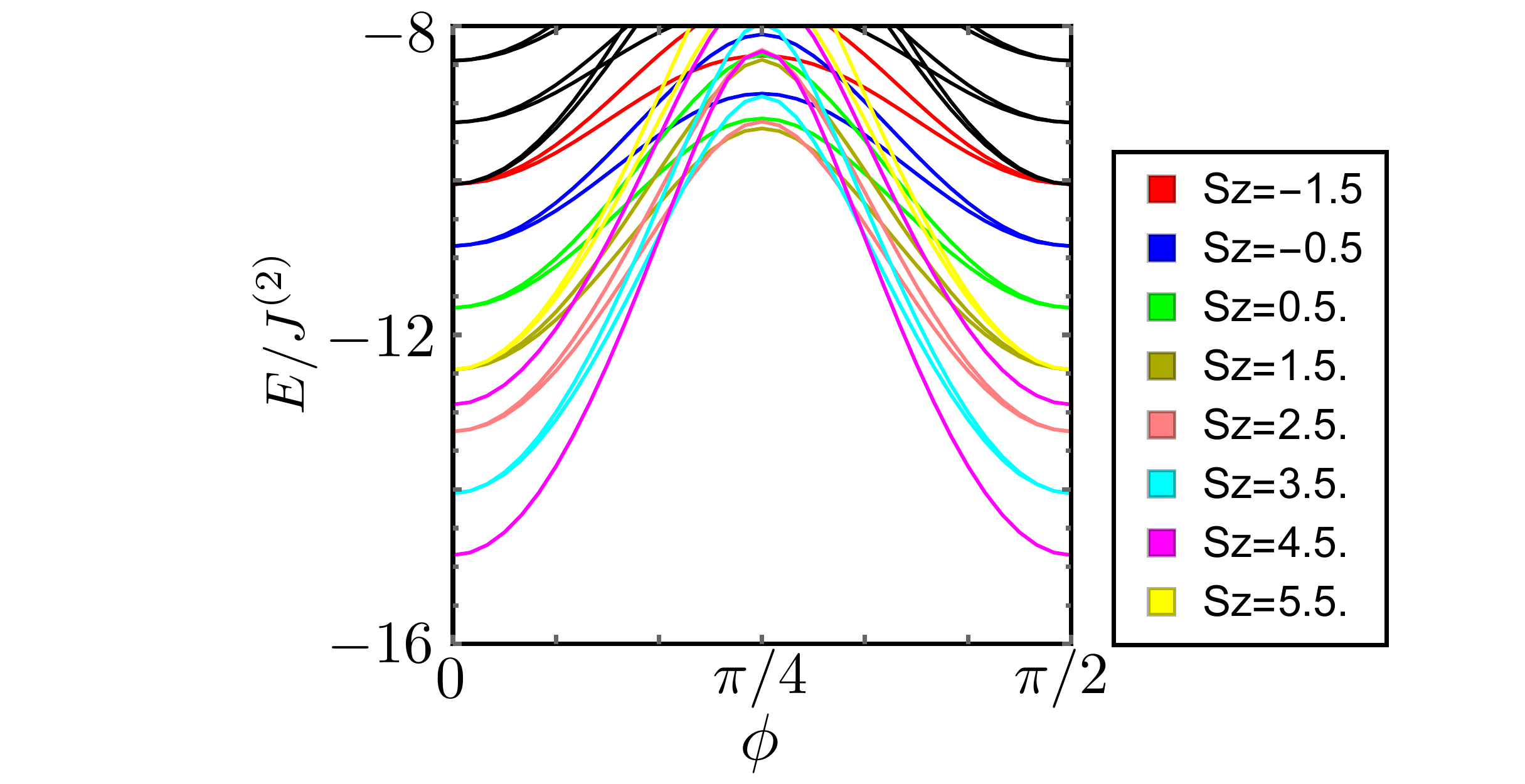}
 \end{center}
 \caption{ Numerical results for 
 the spin model~\eqref{Eq: Hspin}.
 Energy spectra $E/J^{(2)}$ as functions of $\phi$. Only four lowest energies
 in the subspace specified by $S^z_\text{tot}$ are shown here.
 The data are obtained for $N_\text{UC}=3\times3$ at $J^{(3)}/J^{(2)}=0.1$
 under the PBC. The black lines express the energy for 
 $S^z_\text{tot}<-1.5$ or $5.5<S^z_\text{tot}$.}
 \label{fig:SPIN_PERIO_SUP}
\end{figure}
In this appendix, we provide numerical evidences that the quantum 
phase with $S^z_\text{tot}=N_\text{UC}/2$ shown in Fig.~\ref{fig:SPIN_PERIO} 
(a) is gapped.
In Fig.~\ref{fig:SPIN_PERIO_SUP}, we show the $\phi$ dependence of the energy
$E/J^{(2)}$ for $J^{(3)}/J^{(2)}=0.1$ under the PBC
with $N_\text{UC}=3\times3$. This figure clearly demonstrates that 
the ground state with $S^z_\text{tot}=N_\text{UC}/2=4.5$ exhibited in 
$0\leq\phi\lesssim\pi/4$ ($\pi/4\lesssim\phi\leq\pi/2$)
is adiabatically connected to the one of the decoupled system at $\phi=0$ 
($\pi/2$). Thus, one expects that the quantum phase with 
$S^z_\text{tot}=N_\text{UC}/2$ shown in Fig.~\ref{fig:SPIN_PERIO}(a) is 
gapped.

\section{Total $S^z$ of ground state}
\label{Sec:spin_Sz}
\begin{figure}[t]
 \begin{center}
  \includegraphics[width=\columnwidth]{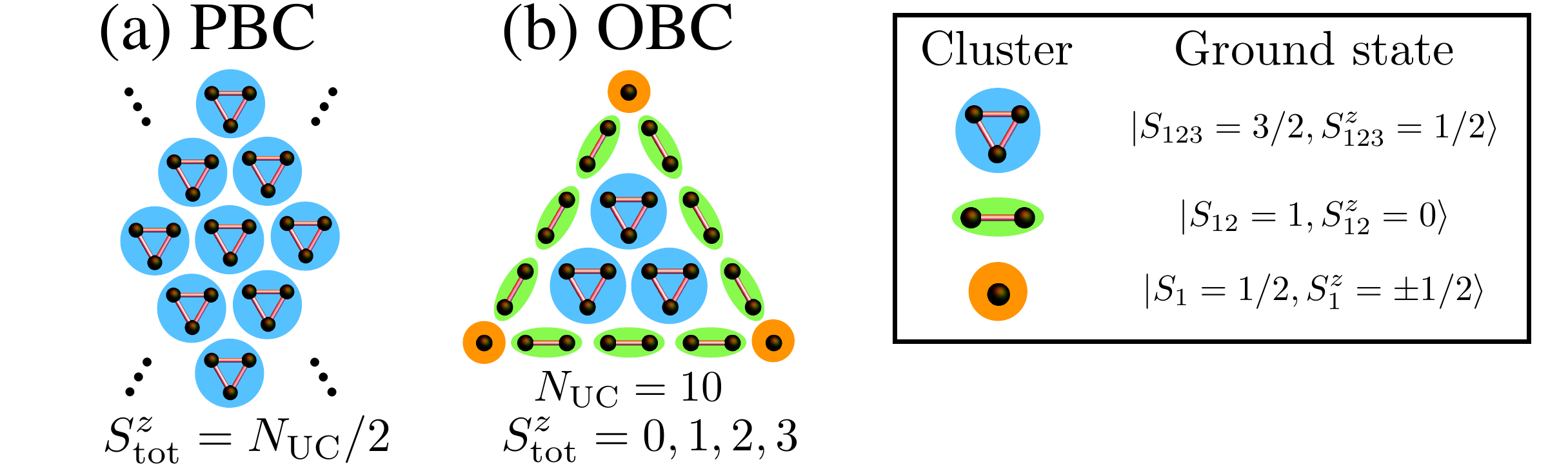}
 \end{center}
 \caption{Effective spin systems at $\phi=0$ under (a) the
 PBC and (b) the OBC. The color expresses types of clusters.
 }
 \label{fig:SPIN_Sz_SUP}
\end{figure}

The periodic system at $\phi=0$ shown in Fig~\ref{fig:SPIN_Sz_SUP}(a)
gives the ground state with $S_\text{tot}^z=N_\text{UC}/2$ as mentioned in
the main text. However, the system under the OBC does not due to boundary 
effects, see Fig~\ref{fig:SPIN_Sz_SUP}(b).
In this appendix, we investigate how 
$S_\text{tot}^z$ of the ground state under the OBC is determined.

Let us first focus on the effective spin model for $N_\text{UC}=10$ as 
discussed in Fig.~\ref{fig:SPIN_OPEN}(a). When setting $\phi=0$, the system
is reduced to three-site, two-site, and one-site problems shown in 
Fig.~\ref{fig:SPIN_Sz_SUP}(b). The total $S^z$ of ground state of each cluster 
is given as $S_{123}^z=1/2$ (trimer), $S_{12}^z=0$ (dimer) and 
$S_{1}^z=\pm1/2$ (isolated site), respectively. Therefore, the 
whole system
gives the eight-fold degenerated ground state whose $S^z_\text{tot}$ ranges 
between 0 and 3, which is obtained by the straightforward calculation 
$(1/2)\times3+0\times9+(\pm1/2)\times3$ [see Fig.~\ref{fig:SPIN_Sz_SUP}(b)].
This is consistent with the results in Fig.~\ref{fig:SPIN_OPEN}(a) at $\phi=0$.

The total $S^z$ of ground states of the cluster for the Hubbard model is
same as that of the spin model. For example, the system for 
$N_\text{UC}=6$ as discussed in Fig.~\ref{fig:HUBBARD}(b) but for $\phi=0$ 
is composed of
one trimer, six dimers and three isolated sites. 
They can be counted by considering the figure similar to 
Fig.~\ref{fig:SPIN_Sz_SUP}(b) but for $N_\text{UC}=6$.
Then, the 
whole system gives the eight-fold degenerated ground state whose magnetization 
ranges between -1 and 2, which is obtained by the straightforward 
calculation $(1/2)\times1+0\times6+(\pm1/2)\times3$.

\section{Weak correlation regime and finite size effect}
\label{Sec:Twist}
\begin{figure}[t]
 \begin{center}
  \includegraphics[width=\columnwidth]{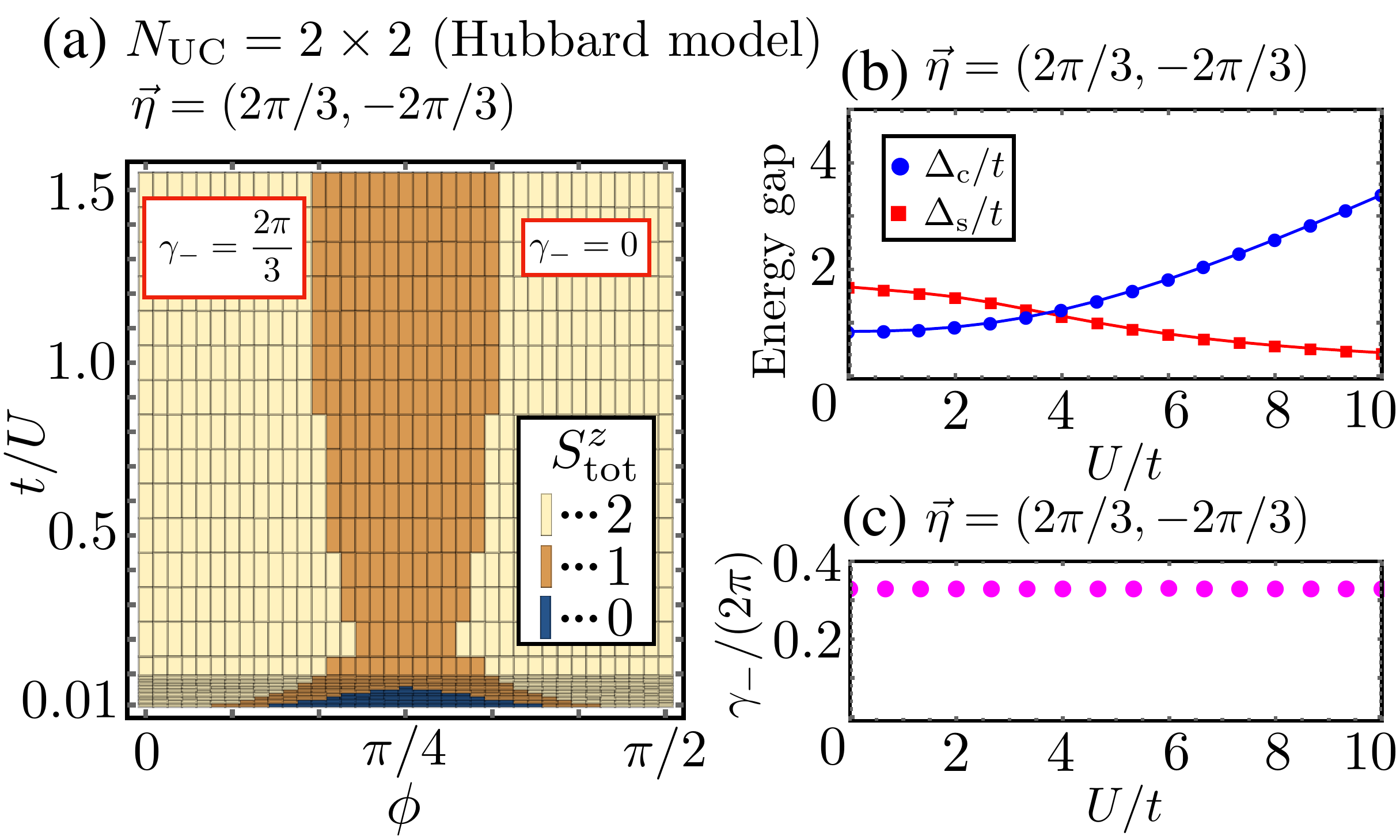}
 \end{center}
 \caption{The same as Figs.~\ref{fig:HUBBARD}(a-c) in the main text but
 for $\vec{\eta}=(2\pi/3,-2\pi/3)$.}
 \label{fig:HUBBARD_SUP}
\end{figure} 
In this appendix, we discuss the finite size effects on the results in 
Fig.~\ref{fig:HUBBARD}(a). To investigate it,
let us introduce the twisted boundary condition in both $x$ and $y$ 
directions. 
We generate 
Fig.~\ref{fig:HUBBARD_SUP}(a) that is the same as Fig.~\ref{fig:HUBBARD}(a)
but for $\vec{\eta}=(2\pi/3,-2\pi/3)$, where $\vec{\eta}=(\eta_x,\eta_y)$ is
the twisted angles.
The figure clearly demonstrates the emergence of the region $S^z_\text{tot}=1$
for the weak correlation regime, which does not exist in 
Fig.~\ref{fig:HUBBARD}(a).
This difference between the two figures suggests that the system size is too 
small to capture the metallic phase that is observed in the noninteracting 
system around $\phi\approx\pi/4$.
On the other hand, the values of $S^z_\text{tot}$ for 
$t_\bigtriangleup/t_\bigtriangledown\lesssim1/2$ and 
$2\lesssim t_\bigtriangleup/t_\bigtriangledown$ shown in both 
Fig.~\ref{fig:HUBBARD}(a) and Fig.~\ref{fig:HUBBARD_SUP}(a) are consistent
with the result of the gapped noninteracting phases. To double check that the 
our conclusion in the main text is not dependent on $\vec{\eta}$, we
here include Figs.~\ref{fig:HUBBARD_SUP}(b) and (c) under 
$\vec{\eta}=(2\pi/3,-2\pi/3)$ that correspond to Figs.~\ref{fig:HUBBARD}(b) and
(c) in the main text. They clearly show the identical behavior.

\section{Eight-fold degeneracy for Hubbard model}
\label{Sec:8fold}
\begin{figure}[t]
 \begin{center}
  \includegraphics[width=\columnwidth]{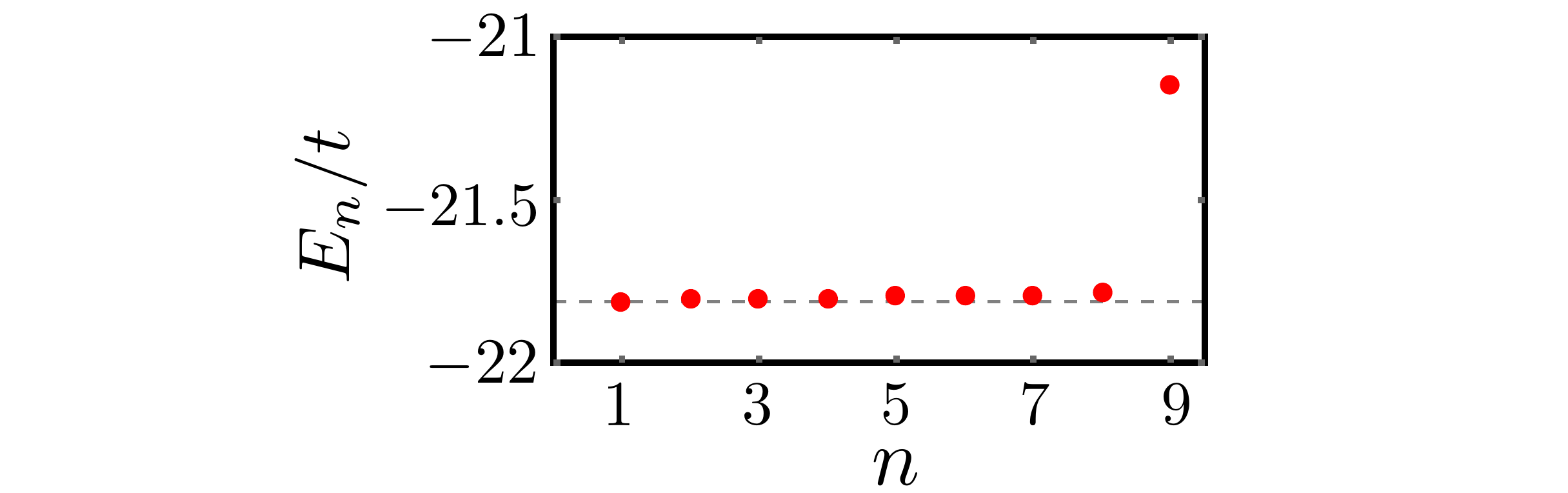}
 \end{center}
 \caption{
 Energy eigenvalues $E_n/t$. We set $U/t=3$, 
 $t_\bigtriangleup/t_\bigtriangledown=0.4$, $N_\text{UC}=6$ and 
 $N_e=3\times N_\text{UC}=18$. 
 The dashed line represents the value of the lowest energy.
 }
 \label{fig:8fold}
\end{figure}

In this appendix, we provide numerical evidences that the ground state 
of the Hubbard model under the OBC shows the eight-fold degeneracy as is the 
case of
the effective spin model calculation in Fig.~\ref{fig:SPIN_OPEN}. In 
Fig.~\ref{fig:8fold}, we show the energy $E_n/t$ for $U/t=3$ and 
$t_\bigtriangleup/t_\bigtriangledown=0.4$ under the OBC with $N_\text{UC}=6$.
It demonstrates the eight-fold degeneracy of the ground state, which
implies the emergence of the corner-Mott states.

\end{document}